\documentclass{article}
\usepackage[dvipdfmx]{graphicx}
\usepackage{grffile}
\usepackage{amsmath}
\usepackage{bm}
\usepackage{cite}
\usepackage{url}
\usepackage{authblk}
\title{Structural balance of alliance and rivalry networks in international relations}
\author[1,2]{Koji Oishi\thanks{oishi@sipeb.aoyama.ac.jp}}
\author[1]{Kentaro Sakuwa\thanks{ksakuwa@aoyamagakuin.jp}}
\affil[1]{\small Department of International Politics, Aoyama Gakuin University, 4-4-25 Shibuya, Shibuya-ku, Tokyo 150-8366, Japan.}
\affil[2]{\small Japan Society for the Promotion of Science, 5-3-1 Kojimachi, Chiyoda-ku, Tokyo 102-8471, Japan.}
\date{\today}
\date{}
\begin{document}

\maketitle
\begin{abstract}
  Does the enemy of my enemy become my friend?
  A growing literature on structural analysis of interstate relationships has tackled this old question from the network perspective.
  However, the mechanism of long-term change in the structure of cooperation and enmity has yet to be fully understood.
  In search for a general explanation for the long-term evolution of interstate structure, we empirically examine the structural balance theory which predicts that a signed network evolves toward a more ``balanced" structure where in many triangular relationships (i.e., triads) two states tend to share a common enemy or three states cooperate with each other. 
  We investigate the network of alliances (positive edges) and rivalries (negative edges) between sovereign states and examine whether its evolution from 1816 to 2009 can be explained by the structural balance theory.
  We find the consistency with the structural balance theory drastically changes over time.
  The empirical pattern follows the prediction by the theory before the German unification in the nineteenth century and after World War II while inconsistent in the middle period.
  This result reveals the impact of the two historical events on the underlying mechanism of network evolution.
  Moreover, the contrast with previous studies of signed social networks that generally support the structural balance theory indicates that international alliance and rivalry networks can be a promising material to study novel mechanisms behind the time evolution of signed networks.
\end{abstract}

\section*{Introduction}
How does the structure of international cooperation and conflict change over time?
Does the international network evolves in a way that improves structural balance?
Using the data of international alliances and rivalries from 1816 to 2009, we empirically examine whether the signed network of international cooperation and enmity and its evolution are characterized by the structural balance hypothesis.

Network representation is a powerful tool to study complex social systems.
When we describe social systems with networks, it is common that edges represent positive relationships such as friendship or collaboration between actors.
However, negative relationships, such as conflict, dispute, and hostility, are also essential parts of social relations, especially international relations.
Signed networks is thus an important field of study of complex social systems \cite{Harrigan2020}.

International relations is a typical example of signed networks as some sovereign states (i.e., countries) are friends while others are foes.
Moreover, they are time-evolving networks because international relations quickly turn from cooperative to antagonistic and vice versa.
Though such changes of their relations have significant impacts on international security, underlying mechanisms of the dynamics have not been fully understood.

\begin{figure}
\centering
\includegraphics[width = 12cm]{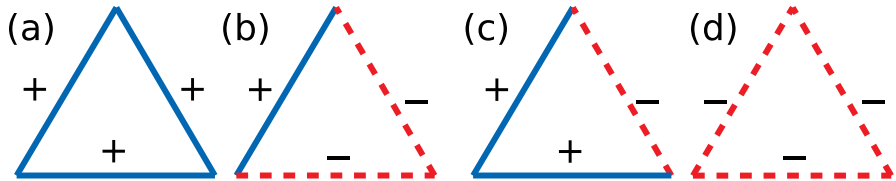}
\caption{(a), (b) Balanced ($+++$ and $+--$) triads. (c), (d) Imbalanced ($++-$ and $---$) triads.}
\label{fig:example}
\end{figure}

A well-known theory for the evolution of signed networks is Heider's structural balance theory, which was originally proposed in social psychology and later formulated in network models (e.g., \cite{Antal2005}).
According to the structural balance theory, signed social networks evolve to increase balanced triads, in which either the enemy of your enemy or the friend of your friend is your friend.
In other words, triads with zero or two negative edges are balanced (Figs. \ref{fig:example}(a) and \ref{fig:example}(b)), while those with one or three negative edges are imbalanced (Figs. \ref{fig:example}(c) and \ref{fig:example}(d)), and balanced triads are expected to be more stable than imbalanced ones.

The structural balance theory has not been fully tested in the long-term evolution of interstate networks, while the theory is supported in various studies of social networks of individuals and animals \cite{Leskovec2010,Szell2010b,Ilany2013}.
Only a few studies have measured the structural balance in international relations \cite{Li2017,Aref2019,Kirkley2019}, which mainly focus the post-World War II period.
In this article, we examine a much longer period (1816--2009) because we are interested in fundamental properties of international relations as time-evolving signed networks rather than specific historical periods or events.
Moreover, we define negative edges between states by interstate rivalries in contrast with the previous studies in which negative edges represent military disputes or war occurrences \cite{Li2017,Aref2019,Kirkley2019,Maoz2007,Warren2010}.
Military disputes are neither necessary conditions for nor an appropriate index of a sustained antagonistic relationship between sovereign states.
For example, the United States and the Soviet Union were antagonistic during the Cold War even when they did not directly engage in direct military clashes. Military clashes are short-term events and states may or may not engage in overt military conflict events all the time even if they are involved in a long-term sustain antagonistic relationships. 
Thus the occurrence of military dispute events are not an ideal measurement of hostile relationships among nations, especially for studying the long-term network evolution.
Therefore, we test whether the structural balance theory holds in the evolution of signed networks of international relations over a longer period, using a better measurement than existing studies.

\section*{Methods and Materials}
We construct the network of alliances and rivalries between sovereign states for each year between 1816 and 2009.
The networks are non-directed unweighted.
Nodes are sovereign states that existed in the year.
We refer to the Correlates of War Project, which lists sovereign states from 1816 to 2016, to identify sovereign states \cite{CorrelatesofWarProject2017}.
Two nodes are connected via positive or negative edges if corresponding two states jointly participated in at least one military alliance or engaged in an interstate rivalry in the year, respectively.
Following previous studies \cite{Li2017,Kirkley2019}, they are connected with a negative edge when the two states have both an alliance and a rivalry.

We draw on the Alliance Treaty Obligations and Provisions (ATOP) dataset \cite{Leeds2002} to identify military alliances.
Though the ATOP dataset collects information of various types of formal military alliances and treaties between states from 1815 to 2016, we use only offensive and defensive alliances to measure positive edges because others (e.g., neutral and non-aggression treaties) do not always indicate the cooperative relation between signatory states are strong enough.
To measure interstate rivalries, we refer to Dreyer and Thompson (2011) \cite{DavidDreyer2011}, who identify pairs of states that regards each other as an enemy based on various sources including official documents of the governments.

\begin{figure}
\centering
\includegraphics[width = 6cm]{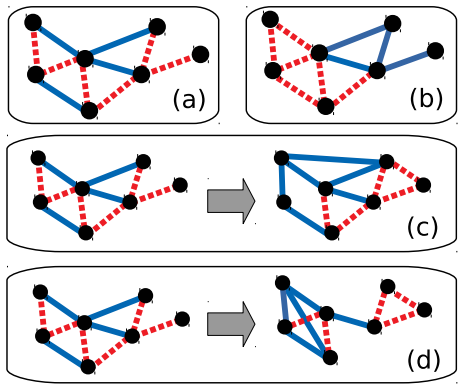}
\caption{Example of randomization of a signed network and its changes.
  Blue solid lines are positive edges and red dashed lines are negative edges.
  The network in (a) is an example of surrogates of the network in (b).
  Both have the same topology and same number of positive and negative edges (four and six, respectively).
  The network changes in (c) is an example of surrogates of the network changes in (d).
  Both has the same numbers of edge state transitions: one negative edge becomes absent, one positive edge becomes absent, one absent edge becomes positive, one absent edge becomes negative, and one negative edge becomes positive.
}
\label{fig:surrogate}
\end{figure}
We compare the empirical network in each year with surrogate networks obtained by shuffling the signs of edges (Figs. \ref{fig:surrogate}(a) and \ref{fig:surrogate}(b)).
Connections between nodes and the numbers of the positive edges and negative edges are held identical between surrogate networks and the empirical one.
The comparison reveals whether the structural balance in a certain year is explained only by the topology and the fraction of edge signs of the network in the year.
The structural balance theory expects that the empirical fraction of balanced triads is significantly larger than the surrogate.

We also compare actual changes of the network in each year (empirical growth) with randomized changes of the network, i.e., surrogate growth (Figs. \ref{fig:surrogate}(c) and \ref{fig:surrogate}(d)).
On the one hand, when we randomize the changes of the network, we conserve the numbers of edge state transitions between three states: positive, negative, and absent.
On the other hand, we randomly choose edges to which the edge transitions realize to.
Therefore, the comparison reveals whether a given property of one-year growth of the network is simply explained by the frequency of edge state transitions in the year.

Specifically, we compare triad state transition in the empirical and surrogate growths of the networks.
Triads are either imbalanced, open, or balanced (denoted hereafter by $-1, 0, +1$, respectively).
Open triads have at least one pair of nodes not connected, so they are neither balanced nor imbalanced.
The triad transition probability $w_{i \to j}$ in year $t$ is the probability that a triad in state $i$ in year $t-1$ takes state $j$ in year $t$, which satisfies $\sum_{j}w_{i \to j} = 1$.

Following the structural balance theory, we expect that: (1) imbalanced triads are not stable and tend to change to open or balanced triads; (2) when open triads become closed, they tend to be balanced rather than imbalanced; (3) balanced triads are stable so tend to stay balanced without changing to either open or imbalanced. Therefore, we expect: (1) $w_{-1 \to +1} + w_{-1 \to 0}$; (2) $w_{0 \to +1} - w_{0 \to -1}$; (3) $w_{+1 \to +1}$ are larger in empirical data compared with the surrogates.
Note that we take the sub-networks of sovereign states which existed in both year $t-1$ and $t$ to measure edge and triad state transitions and generate surrogates growth, because in this study we assume that emergence and demise of sovereign states are out of the scope of the structural balance theory.

We employ z-score to evaluate the difference between empirical and surrogate networks.
The z-score of a variable is
\begin{equation}
  z = \frac{{x} - \mu}{\sigma},
\end{equation}
where $x$ is the empirical value of the variable, $\mu$ and $\sigma$ are the mean and the standard deviation of the variable in surrogates.
$|z| > 2$ is a benchmark for a significant difference between empirical values and surrogate values, as $z > 2$ ($z < 2$) means the empirical value is significantly larger (smaller) than surrogates.

\section*{Results}
\begin{figure}
\centering
\includegraphics[width = 12cm]{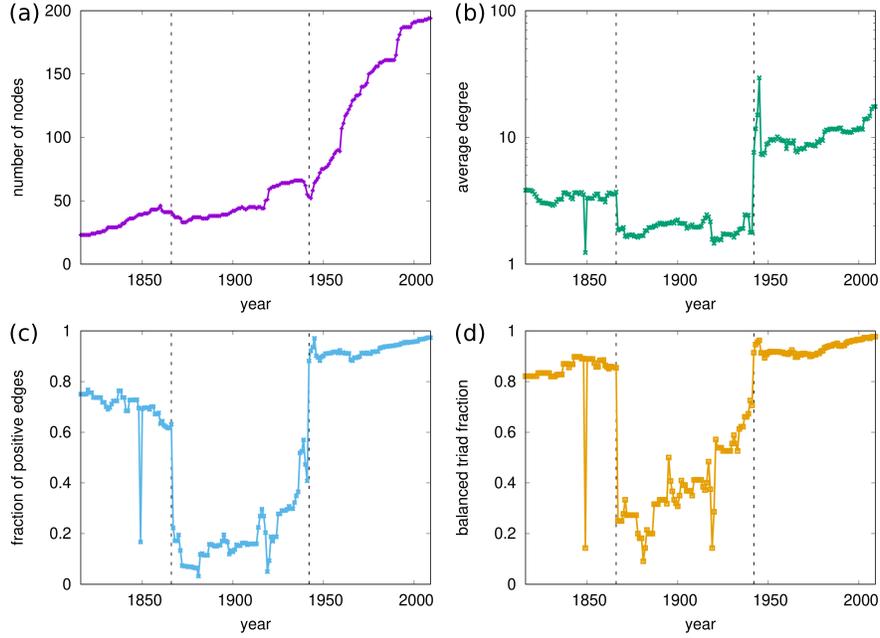}
\caption{
  (a) The number of nodes, (b) the average degree, (c) the fraction of positive edges and (d) the fraction of balanced triads.
  Vertical dashed lines indicate the end of the first period (1866) and the beginning of the last period (1942).
}
\label{fig:character}
\end{figure}

First we find the static properties of the alliance and rivalry network are clearly different across three periods, (1) 1816--1866, (2) 1867--1941, and (3) 1942--2009 (Fig. \ref{fig:character}).
While the number of nodes (sovereign states) tends to increase over all the periods (Fig. \ref{fig:character}(a)), the average degree drops in 1867 and jumps up in 1942 (Fig. \ref{fig:character}(b)).
By the same token, the fraction of positive edges and the fraction of balanced triads (Figs. \ref{fig:character}(c) and \ref{fig:character}(d), respectively) are constantly high (i.e., positive edges and balanced triads are the majority) during 1816--1866 and 1942--2009, while both fractions suddenly drooped in 1867 and gradually increased between until 1941.
Thus, the balanced triads are not always the majority and their fraction did not always increase.
The pre-1867 and post-1942 periods are quite different than the period in-between in terms of the static nature of structural balance.

\begin{figure}
\centering
\includegraphics[width = 12cm]{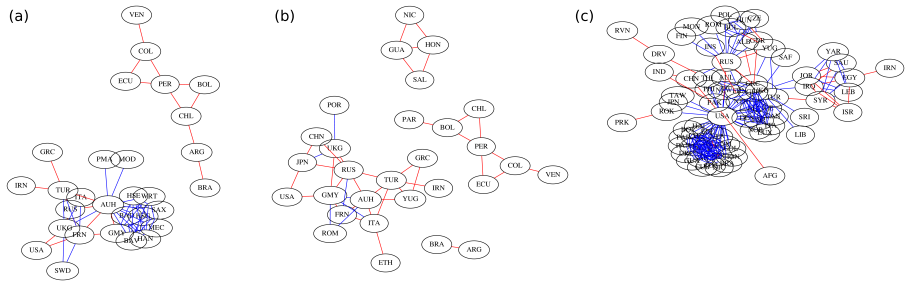}
\caption{
  Alliance and rivalry networks in (a) 1855, (b) 1905, and (c) 1955.
  Node labels are the COW abbreviation of sovereign states' names \cite{CorrelatesofWarProject2017}.
}
\label{fig:vizual_year_period}
\end{figure}

Fig. \ref{fig:vizual_year_period} shows typical networks in each period.
The networks in 1855 and 1955 have clusters which densely connected by positive edges.
The networks in 1955 is larger and have more clusters with positive edges compared with that in 1855.
In contrast, the network in 1905 does not have clusters of positive edges, it is sparser than the other two, and most of its edges are negative.

\begin{figure}[!htb]
\centering
\includegraphics[width = 12cm]{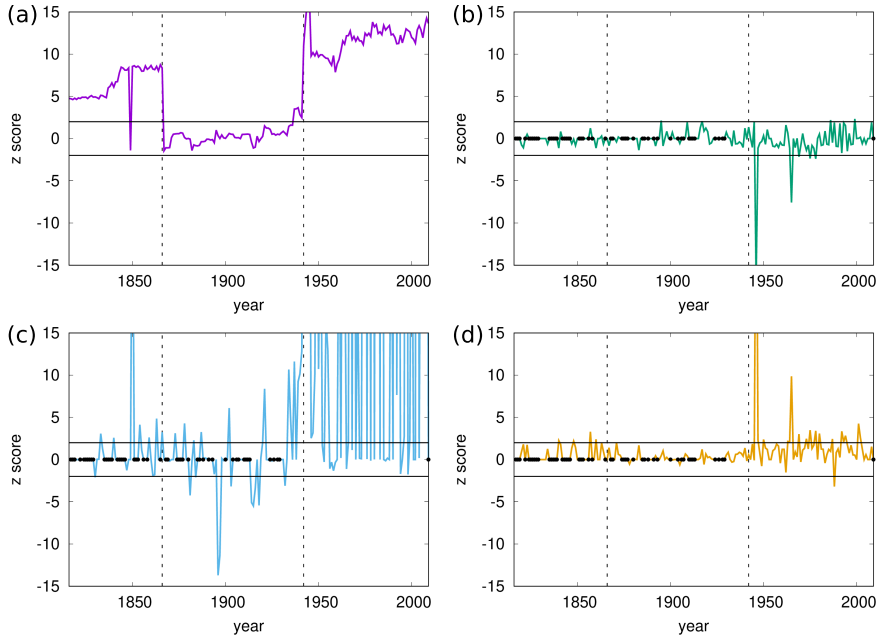}
\caption{
  (a) Z-score of balanced triads between empirical and surrogate networks.
  (b)--(d) Z-score between empirical and surrogate growth of (b) $w_{-1 \to +1} + w_{-1 \to 0}$, (c) $w_{0 \to +1} - w_{0 \to -1}$, and (d) $w_{+1 \to +1}$.
  They are calculated from 1,000 samples.
  Horizontal solid lines indicate $|z| = 2$.
  Vertical dashed lines indicate the end of the first period (1866) and the beginning of the last period (1942).
  Black filled circles indicate the years when no edge state transitions are observed for the focal sub-networks.
}
\label{fig:zscore}
\end{figure}

We further compare the fraction of balanced triads of the empirical network in each year with surrogate networks in which signs of edges are randomly shuffled without changing the topology of the empirical network (Fig. \ref{fig:zscore}(a)).
During the first period (1816--1866) and third period (1942--2009), the fraction of balanced triads is significantly larger than that of surrogate networks ($|z| > 2$), which is consistent with the structural balance theory.
On the other hand, between 1867 and 1941, the difference between the empirical and the surrogate in the second and third periods is much less clear ($|z| < 2$), therefore the support for the structural balance theory is weaker in this period.

Next, we compare the growth of the networks in each year and its surrogate growth for the year.
Figs. \ref{fig:zscore}(b)--(d) show z-scores of the three values, (b) $w_{-1 \to +1} + w_{-1 \to 0}$, (c) $w_{0 \to +1} - w_{0 \to -1}$, and (d) $w_{+1 \to +1}$, to compare empirical and surrogate growth.
The structural balance theory posits that these values should be larger in empirical data than in surrogates.
Note that when no edge state transitions are observed in the focal sub-network in the empirical date, we cannot generate surrogates because there is nothing to randomize.
Such years are indicated with black circles in Figs. \ref{fig:zscore}(b)--(d).

We again see that the support for the structural balance theory is different over time.
Open triads were likely to become balanced rather than imbalanced in the early and especially the recent periods (Fig. \ref{fig:zscore}(c)).
On the other hand, such tendency is not clear in the middle period as z-score only sometimes satisfies $|z| > 2$ and both $z > 2$ and $z < -2$ are observed with a similar frequency. 
Balanced triads tended to stay balanced especially in the postwar period, but not quite in earlier periods.
Z scores larger than 2 are observed only sometimes before the 1940's (Fig. \ref{fig:zscore}(d)).
There is no concrete evidence that imbalanced triads were likely to become either open or balanced as the theory predicts---$|z|$ hardly exceeded 2 except for a few years in the recent period (Fig. \ref{fig:zscore}(b)).
Taken together, the comparison of empirical triad state transitions with surrogate growth expected from edge state transition supports the theory in the early period and, more clearly, in the recent periods, while not in the middle period.

\section*{Discussion}
The changes of consistency with the structural balance theory coincided with two important historical events: German unification in the nineteenth century and World War II.
This observation partly explains the abrupt changes of network characteristics.
The ``middle" period in our analysis (1867--1941) starts with German unification and ends during World War II.

Shortly prior to German unification, Prussia defeated Austria in 1866 (Austro-Prussian War) and annexed several small states such as Saxony and Hanover, which had densely allied each other before the unification.
In the language of networks, nodes connected via dense positive edges and almost no negative edges (therefore forming largely balanced triads) merged into one node. It resulted in abrupt reduction in the fraction of positive edges and balanced triads.
In the interwar period and during World War II, both the Axis powers and the Allies increased their alliances. The Allies defeated the Axis powers and ``absorbed" them (West Germany, Japan, and Italy) into the large Western alliance.
The fraction of positive edges and balanced triads increased first, in the interwar period, by the formation of dense alliances within the two camps, and later in the postwar period by the elimination of rivalries between the camps due to the absorption of one camp by the other.

Moreover, our data analysis demonstrated that German unification and World War II changed the dynamics of network evolution rather than merely shifted network characteristics in the specific years.
The period between these events was quite distinct in that triadic interactions, i.e., the tendency toward balanced triads are not observed beyond the expectation from random addition and deletion of edges.
This implies that international relations during the period was mostly driven by dyadic motivations and interactions in which pairs of states cooperate or confront over bilateral issues, while triadic (or poly-adic) considerations on structural balance involving third parties did not strongly drive international politics.
On the other hand, changes of alliances and rivalries between two states were strongly influenced by considerations to enhance balance in their relationships with others as the structural balance theory predicts before German unification or after World War II.
Network analysis thus provides this new insight into the impacts of the two historical events.

We make a contribution to network science by finding that international alliance and rivalry networks can be a promising material to find novel dynamics of signed networks.
In contrast with our result, previous studies on signed networks of individuals or animals are generally supportive to the structural balance theory \cite{Leskovec2010,Szell2010b,Ilany2013}.
It can be partly because nodes do not merge by definition in their case, while the merger of nodes seems to play an important role in abruptly changing the consistency with the structural balance theory in our case.
Existing models of the structural balance theory also fixes the set of nodes \cite{Antal2005}, while the split and merger of social groups, e.g., firms and political parties, are common phenomena.
Therefore, we conjecture interaction between structural balance and merger of nodes into models can help to reproduce the long-time evolution of signed networks, including the alliance and rivalry networks.

Note that some previous studies on structural balance of interstate relations take longer cycles into account when measuring structural balance of signed networks \cite{Li2017,Kirkley2019}, e.g., whether the enemy of the enemy of your enemy is an enemy or not.
Though the current study focused on the structural balance of triads, the long-time development of structural balance in longer cycles is also an interesting direction to expand.

\section*{Availability of data and materials}
The sovereign states membership data is available at\\\url{https://correlatesofwar.org/data-sets/state-system-membership}.
The interstate alliance data is available at \url{http://www.atopdata.org/}.
The interstate rivalry data is taken from Dreyer and Thompson (2011) \cite{DavidDreyer2011}.

\section*{Competing interests}
The authors declare that they have no competing interests.
\section*{Funding}
This research is supported by JSPS KAKENHI Grant Number JP20J01060 (K.O.) and JP20K13431 (K.S.).
The funders had no role in study design, data collection and analysis, decision to publish, or preparation of the manuscript.
\section*{Author's contributions}
K.O. and K.S. conceived and designed the research; K.O. conducted the analyses; K.O. and K.S. wrote the paper.
\section*{Acknowledgements}
Not applicable.

\begin{flushleft}

\end{flushleft}


\begin{thebibliography}{10}

\bibitem{Harrigan2020}
Nicholas~M. Harrigan, Giuseppe~(Joe) Labianca, and Filip Agneessens.
\newblock {Negative ties and signed graphs research: Stimulating research on
  dissociative forces in social networks}.
\newblock {\em Soc. Networks}, 60(October 2019):1--10, 2020.

\bibitem{Antal2005}
T.~Antal, P.~L. Krapivsky, and S.~Redner.
\newblock {Dynamics of social balance on networks}.
\newblock {\em Phys. Rev. E - Stat. Nonlinear, Soft Matter Phys.}, 72(3):1--10,
  2005.

\bibitem{Leskovec2010}
Jure Leskovec, Daniel Huttenlocher, and Jon Kleinberg.
\newblock {Signed networks in social media}.
\newblock {\em Conf. Hum. Factors Comput. Syst. - Proc.}, 2:1361--1370, 2010.

\bibitem{Szell2010b}
Michael Szell, Renaud Lambiotte, and Stefan Thurner.
\newblock {Multirelational organization of large-scale social networks in an
  online world}.
\newblock {\em Proc. Natl. Acad. Sci. U. S. A.}, 107(31):13636--13641, 2010.

\bibitem{Ilany2013}
Amiyaal Ilany, Adi Barocas, Lee Koren, Michael Kam, and Eli Geffen.
\newblock {Structural balance in the social networks of a wild mammal}.
\newblock {\em Anim. Behav.}, 85(6):1397--1405, 2013.

\bibitem{Li2017}
Weihua Li, Aisha~E. Bradshaw, Caitlin~B. Clary, and Skyler~J. Cranmer.
\newblock {A three-degree horizon of peace in the military alliance network}.
\newblock {\em Sci. Adv.}, 3(3):1--11, 2017.

\bibitem{Aref2019}
Samin Aref, Mark~C. Wilson, and Ernesto Estrada.
\newblock {Balance and frustration in signed networks}.
\newblock {\em J. Complex Networks}, 7(2):163--189, 2019.

\bibitem{Kirkley2019}
Alec Kirkley, George~T. Cantwell, and M.~E.J. Newman.
\newblock {Balance in signed networks}.
\newblock {\em Phys. Rev. E}, 99(1):1--11, 2019.

\bibitem{Maoz2007}
Zeev Maoz, Lesley~G. Terris, Ranan~D. Kuperman, and Ilan Talmud.
\newblock {What is the enemy of my enemy? Causes and consequences of imbalanced
  International relations, 1816-2001}.
\newblock {\em J. Polit.}, 69(1):100--115, 2007.

\bibitem{Warren2010}
T.~Camber Warren.
\newblock {The geometry of security: Modeling interstate alliances as evolving
  networks}.
\newblock {\em J. Peace Res.}, 47(6):697--709, 2010.

\bibitem{CorrelatesofWarProject2017}
{Correlates of War Project}.
\newblock {State System Membership List, v2016}, 2017.

\bibitem{Leeds2002}
Brett~Ashley Leeds, Jeffrey~M. Ritter, Sara Mc~Laughlin Mitchell, and Andrew~G.
  Long.
\newblock {Alliance treaty obligations and provisions, 1815-1944}.
\newblock {\em Int. Interact.}, 28(3):237--260, 2002.

\bibitem{DavidDreyer2011}
David Dreyer and William~R. Thompson.
\newblock {\em {Handbook of International Rivalries}}.
\newblock CQ Press, Washington, DC, 2011.

\end{thebibliography}
\end{document}